# Suppression of radiation loss by hybridization effect in two coupled split-ring resonators

T. Q. Li[1], H. Liu[1, *], T. Li[1], S. M. Wang[1], J. X. Cao[1], Z. H. Zhu[1], Z. G. Dong[1], S. N. Zhu[1] and X. Zhang[2]

[1]*Department of Physics, National Laboratory of Solid State Microstructures, Nanjing University, Nanjing 210093, People's Republic of China*
[2]*5130 Etcheverry Hall, Nanoscale Science and Engineering Center, University of California, Berkeley, California 94720-1740, USA*

Email address: liuhui@nju.edu.cn.
URL: http://dsl.nju.edu.cn/dslweb/images/plasmonics-MPP.htm

Abstract

This paper investigates the radiation properties of two coupled split-ring resonators (SRRs). Due to electromagnetic coupling, two hybrid magnetic plasmon modes were induced in the structure. Our calculations show that the radiation loss of the structure was greatly suppressed by the hybridization effect. This led to a remarkable increase in the Q-factor of the coupled system compared to the single SRR. By adjusting the distance between the two SRRs, the Q-factor changed correspondingly due to different electromagnetic coupling strengths. This resulted in a coupled structure that functioned as a new type of nanocavity with an adjustable Q-factor.

Research on magnetic metamaterial has attracted significant interest ever since Pendry first reported that split-ring resonators (SRR), otherwise known as nonmagnetic metallic structures with sizes below the diffraction limit, exhibit negative permeability [1]. Other studies have likewise shown that the effective medium composed of SRRs can support resonant magnetic plasmon (MP) oscillation analogous to surface plasmon resonance [2-9]. Combined with a material displaying the characteristic of negative permittivity at the same response frequency region, a negative refraction effect has been produced [10], leading to a wealth of research into metamaterials. In addition to negative refraction, Pendry also put forward the proposal that SRRs could be used to enhance nonlinear optical phenomena [1]. In recent years, some groups have begun to apply magnetic resonance nanostructures to nonlinear optical effects such as SHG [11], nanolaser [12], and SPASER [13]. Given that all of these nonlinear processes are based on resonance behavior, a high quality factor (Q-factor) has been pursued to improve the efficiency of the magnetic plasmon structures. Various methods have been introduced to improve the resonance properties of magnetic structures including weak asymmetric structures that enable the excitation of trapped modes [14]. However, due to dramatic radiation loss caused by strong coupling to free space, the Q-factor of the magnetic resonators is still quite low [15]. Accordingly, the inhibition of radiation loss is necessary in obtaining high Q magnetic nanocavities.

Although metamaterials have many interesting applications, the coupling interactions between the elements in such metamaterials are somewhat ignored by most of researchers in this field, given that the effective properties of metamaterials can be viewed as the "averaged effect" of the resonance property of individual elements. However, the coupling interaction between elements should always exist when they are arranged together into real practical metamaterials. Sometimes, especially when the elements are very close, this coupling effect is not negligible and will have a substantial effect on the metamaterial's properties. In recent years, the coupling effect in such materials has aroused more attention from researchers undertaking studies on topics such as pairs of split-ring resonators [16-18], pairs of nano-rods [19-20], magnetic wave along meta-elements [21-24], and other interesting coupled metamaterials [25]. The possible related applications have also been reported by several groups.

Recently, our group studied the resonance properties of two coupled SRRs, which can be called as the magnetic dimer (MD). Strong magnetic coupling interaction is established between the two SRRs, through which we introduced two hybrid magnetic plasmon modes: a lower frequency bonding mode and a higher frequency antibonding mode [26]. Optical activity resulting from the hybridization effect was experimentally observed in this MD system [27].Moreover, this type of MD structure has also been recently reported to construct stereometamaterials with different twisted angles [28].

In this paper, we analyzed the radiation properties of the coupled MD structure in the THz region for the first time. This region marks the area wherein simultaneous magnetic and electric couplings coexist. By studying the far field radiation of this structure, we found two different radiation patterns in the two hybrid modes; we believe these could be attributed to the different resonant behaviors. Compared with the single split-ring resonator, we found that the radiation loss was greatly suppressed at the lower symmetry mode, resulting in a dramatic increase in the Q-factor of this structure. Moreover, the Q-factor changed continuously together with the variations that occurred in the distance between the two SRRs. This leads to a possible design for a nanocavity with an adjustable Q-factor.

Figure 1 illustrates the geometry of the MD metamaterials along with their design parameters. Each unit cell consisted of two spatially separated identical SRRs, twisted at an angle of $\varphi = 180^0$ with respect to one another. The MD was made of gold and was embedded in a homogeneous dielectric with $\varepsilon = 1$ (air).

In order to quantitatively study the resonance behavior of MD, a commercial software package CST MICROWAVE STUDIO (Computer Simulation Technology GmbH, Darmstadt, Germany) was employed to obtain numerical analysis. In the calculations, the permittivity of gold is given by the Drude model, $\varepsilon(\omega) = 1 - \omega_p^2/(\omega^2 + i\omega_\tau \omega)$, where $\omega_p$ is the bulk plasma frequency and $\omega_\tau$ is the relaxation rate. For gold, the characteristic frequencies fitted to experimental data are $\omega_p = 1.37 \times 10^4 \, \text{THz}$ and $\omega_\tau = 40.84 \, \text{THz}$. [29] We used normally incident light with polarization along the y-direction as we carried out the simulations for the excitation of these SRR dimer metamaterials (Fig. 1). Probes were set in the center of two splits to record the local field. In the figure, the positions are shown as red arrows. When the incident frequency was swept from 0 to 120 THz, we obtained two resonance peaks on the local electric field curve (See Fig. 2a) which corresponded with the hybrid magnetic plasmon modes. In order to explain these two resonance peaks, Lagrangian formalism was then introduced to describe the coupled system.

One SRR can be viewed as an equivalent LC circuit, in which the metal ring is regarded as a magnetic loop with inductance L, and the slit of the ring is a capacitor with capacitance C. Thus, this system has a resonance frequency of $\omega_0 = 1/\sqrt{LC}$; the oscillating current induced in the resonator generates the magnetic moment. By defining the charge accumulated in the slit as a generalized coordinate, the Lagrangian of one SRR can be written as $\Im = L\dot{q}^2/2 - q^2/2C$, wherein $L\dot{q}^2/2$ refers to the electrostatic energy stored in the ring and $q^2/2C$ refers to the energy in the slit. Accordingly, with additional magnetic and electric interaction terms, the Lagrangian of this coupled system is composed of a combination of the two individual SRRs [28]:

$$\Im = \frac{L}{2}\left(\dot{q}_1^2 - \omega_0^2 q_1^2\right) + \frac{L}{2}\left(\dot{q}_2^2 - \omega_0^2 q_2^2\right) + K_m \dot{q}_1 \dot{q}_2 + K_E \omega_0^2 q_1 q_2. \tag{1}$$

Here, Q1 and Q2 are oscillating charges in the respective SRRs; and $K_m$ and $K_E$ are the mutual inductances for the magnetic and electric interactions, respectively. By solving the Euler-Lagrange equations, two Eigenfrequencies are obtained as:

$$\omega_- = \omega_0 \sqrt{\frac{1-\kappa_e}{1+\kappa_m}} \quad \text{and} \quad \omega_+ = \omega_0 \sqrt{\frac{1+\kappa_e}{1-\kappa_m}}. \tag{2}$$

Here, $\kappa_m = K_m/L$, $\kappa_e = K_e/L$, both of which depend on the coupling strength of the two SRRs and could be obtained in the following simulations. The bonding mode, $|\omega_-\rangle$, demonstrates the symmetric charge distribution $(q_1 = q_2)$, which is the lower frequency resonant mode. Meanwhile, the antibonding mode, $|\omega_+\rangle$, is characterized by an antisymmetric charge distribution $(q_1 = -q_2)$, which is the higher frequency resonant mode. In our simulation, the three character frequencies $\omega_0$, $\omega_+$, and $\omega_-$ could be obtained under different coupling distances (Tab.1). Correspondingly, the two coupling coefficients, $\kappa_m$ and $\kappa_e$, could be calculated from Eq. (2) (Tab.1). The local magnetic field distributions of the SRRs in the y-z plane and the current density distribution are also given in Fig.2. At the lower frequency resonant mode, the currents of the two SRRs rotated in the same direction (Fig. 2 (d)). Thus, the local magnetic field was enhanced by the summation of the magnetic field generated in each SRR (Fig.2 (b)). At the higher frequency resonant mode, the currents in the two SRRs rotated in opposite directions (Fig. 2 (e)), and the local magnetic field was eliminated by cancelling the magnetic field generated in each SRR (Fig.2 (c)).

By calculating the radiation pattern of this coupled system, different radiation behaviors were observed at the bonding and antibonding modes. Moreover, through the simulation undertaken using a CST package, the radiation skins (radiation power density distribution on the surface of a sphere with radius of one meter) of MD at a specified frequency could be obtained directly. Figure 3 shows the radiation skins of MD at the two resonance hybrid modes. For the purpose of comparison, the radiation skin of the resonance mode of a single SRR is also included. The projected curves of these three 3-D skins in the z-y, z-x, and x-y planes are presented in Fig. 4. We can also see that the radiation pattern of the single SRR is a near-ellipsoid with its maximum value expressed as a circular projection on the x-z plane (dashed lines in Fig.4). The radiation pattern for the bonding mode of the structure at the lower resonance frequency looks like a peanut in the z-y and z-x planes (black lines in Fig.4). At the higher resonance frequency of the antibonding mode, the radiation pattern looks like a doughnut, with a maximum value found in the x-z plane and a minimum value on the y axis.

The differing radiation behaviors of the hybrid modes can be explained using a visual representation of the dipole model. At resonance frequencies, the electromagnetic wave strongly coupled with the SRRs and generated an oscillating current in the structure. As a result, a strong electric field was generated within the split gap, and a magnetic field was induced inside the loop. Thus, the equivalent radiation structure of one SRR can be viewed as an electric dipole in the split gap (in the y direction) and a magnetic dipole in the loop (in the z direction). Accordingly, for a single SRR, the radiation pattern is a combination of an electric dipole and a magnetic dipole with their directions perpendicular to each other.

According to classical electrodynamics theory, the radiation power of an electric dipole is much stronger than that of a magnetic dipole [30], resulting in the latter being dominated by the former. Thus, the direction of the maximum radiation of a single SRR is in the z-x plane. However, the radiation behaviors of the two hybrid modes are quite different for the two coupled SRRs. At the lower frequency of the bonding mode, the system is composed of two magnetic dipoles in the same direction and two electric dipoles in opposite directions (Fig.5 (a)). As the electric dipoles cancel each other, the coupled SRRs can be regarded as a magnetic dipole (Fig.5 (b)). The enhanced radiation of the two parallel magnetic dipoles creates the maximum radiation power pattern in the x-y plane. At the higher frequency of the antibonding mode, the system becomes composed of two magnetic dipoles in opposite directions and two electric dipoles in the same direction (Fig.5 (c)). As the magnetic dipoles cancel each other, the coupled SRRs can thus be regarded as an electric dipole (Fig.5 (d)). Thus, the maximum radiation power pattern is in the z-x plane, resulting in the greatly reduced radiation power in the y direction. Accordingly, it is obvious that the coupled structure behaves like an equivalent magnetic dipole in the bonding mode and an equivalent electric dipole in the antibonding mode.

In the above analysis, it can be seen that the hybridization effect of the coupled SRRs system greatly affected the far-field radiation pattern of the coupled SRR structure. The radiation losses in the bonding and antibonding modes also changed dramatically due to this coupling effect. The distance dependence of the radiation resistance of the two hybrid modes of the structure was also investigated in our simulations by varying the distance between the two SRRs. The radiation resistance is defined as $R = P/I^2$, where the P is the radiation power, and I is the current in the system. The simulation result shows the radiation power density $\beta(\theta,\varphi)$ on the surface of a sphere with a radius of one meter (Fig. 3). The radiation power is obtained through the integral of radiation power density on the surface: $P = \int \beta(\theta,\varphi) d\theta d\varphi$. The simulation also provides the current density j of the structure (Figs. 2 (d) and (e)). Thus, we can calculate the current passes through the structure by integrating the current density on the cross section area: $I = \int \vec{j} \cdot d\vec{s}$. The results of radiation resistance, radiation power, and current with the distance varying from 70nm to 140nm are listed in Tab. 2 while the radiation resistance is presented in Fig. 6. As previously anticipated, the radiation resistance and radiation power of the lower resonant frequency of the coupled SRRs (red line) are smaller than those of a single SRR (dashed line). This is attributed to the weaker radiation strength of the magnetic dipole, indicating that less system energy is radiated into the free space. As distance increases, the decrease in the coupling strength between the SRRs reduces the cancellation of the electric dipoles, causing a corresponding increase in radiation resistance and radiation power. On the other hand, at the higher resonant frequency of the antibonding mode (black line), both the radiation resistance and radiation power are higher than those of the single SRR. This is due to the enhancement of the two electric dipoles. However, since the magnetic dipole is much weaker than the electric dipole, the reduction of the cancellation of the magnetic dipoles makes a smaller contribution to the increase in the radiation resistance and radiation power. This was achieved by increasing the distance between the SRRs. Consequently, the change in the radiation resistance and radiation power of the antibonding mode is slight.

Given that the suppression of the radiation loss of the bonding mode led us to anticipate an increase in the Q-factor, we will discuss the Q-factor of the coupled SRRs in the following section. The life of the photon in the structure is defined as $\tau = \frac{2\pi}{\Delta\omega}$, $\Delta\omega$ is the full-width at half-maximum (FWHM) of the resonant peak (Fig.

2 (a)). On the other hand, the corresponding Q-factor of the resonance can be expressed as $Q = \tau\omega = \frac{2\pi\omega}{\Delta\omega}$, where $\omega$ is the resonant frequency of the resonant peak (Fig. 2 (a)). The obtained result of resonant frequency, FWHM, and Q factor under different coupling distances are listed in Tab.3. In our simulations, the Q-factor of the single SRR is obtained as $Q_S = 164.63$. For an MD structure with $D = 70nm$, the Q-factor of the bonding mode is $Q_L = 376.61$, and the Q-factor of the antibonding mode is $Q_H = 129.16$. We can see that $Q_L$ is approximately 2.3 times $Q_S$, while $Q_H$ is smaller than $Q_S$. Thus, it can be seen that the cancellation of the electric dipole of the bonding mode greatly suppressed the radiation loss and reduced the coupling between the MD and the free space. Thus, the Q-factor was much larger than the Q-factor of the single SRR. For the antibonding mode, the enhancement of the electric dipole increased the radiation loss, leading to a smaller Q-factor.

The dependence relationship between the Q-factor of the MD and the coupling distance is depicted by the curves in Fig.7. As was anticipated by the dipole model, the Q-factor at the bonding mode obviously decreased as the distance between the two SRRs increased. However, for the antibonding mode, no obvious change in the Q-factor could be observed because the variation of radiation from magnetic dipole could almost be ignored. Therefore, according to the above results, we can see that a high Q factor can be obtained at the lower frequency bonding mode. Furthermore, it is likely that this coupled structure can provide a method for designing a nanocavity with an adjustable Q-factor by adjusting the distance between the two SRRs.

In conclusion, by analyzing far field radiation of coupled SRRs, we found that the hybridization effect led to different radiation behaviors in the bonding and antibonding modes. The radiation loss was greatly suppressed at the lower frequency bonding mode and was enhanced at the higher frequency antibonding mode. As the distances between the SRRs decreased, the magnitude of the radiation power of the bonding mode correspondingly decreased due to the greater cancellation of the electric dipoles. Furthermore, this suppression of the radiation loss led to a higher

Q-factor of the system at the bonding mode. These results can be used to improve the nonlinear optical efficiency of metamaterials. The dependence relationship of the Q-factor on the distance between the SRRs was also investigated, and this relationship provides a method for designing a nanocavity with an adjustable Q-factor.

This work is supported by the National Natural Science Foundation of China (No.10604029, No.10704036 and No.10874081), and by the National Key Projects for Basic Researches of China (No.2009CB930501, No.2006CB921804 and No. 2004CB619003).

| D (nm) | 70 | 80 | 90 | 100 | 110 | 120 | 130 | 140 |
|---|---|---|---|---|---|---|---|---|
| $\omega_0$ (THz) | 75.65 | 75.65 | 75.65 | 75.65 | 75.65 | 75.65 | 75.65 | 75.65 |
| $\omega_-$ (THz) | 59.02 | 60.60 | 61.91 | 63.01 | 63.93 | 64.71 | 65.38 | 65.94 |
| $\omega_+$ (THz) | 93.12 | 91.95 | 90.79 | 89.67 | 88.60 | 87.58 | 86.62 | 85.71 |
| $\kappa_e$ | 0.1365 | 0.1422 | 0.1427 | 0.1386 | 0.1303 | 0.1184 | 0.1023 | 0.0826 |
| $\kappa_m$ | 0.3084 | 0.2671 | 0.2348 | 0.2102 | 0.1928 | 0.1816 | 0.1768 | 0.1775 |

Tab.1. The simulated values of the resonant frequency of single SRR ($\omega_0$), the resonant frequencies of bonding mode of MD ($\omega_-$) and antibonding mode of MD ($\omega_+$), coupling parameter ($\kappa_e$ and $\kappa_m$) under different coupling distance between two SRRs.

| D (nm) | 70 | 80 | 90 | 100 | 110 | 120 | 130 | 140 |
|---|---|---|---|---|---|---|---|---|
| $P_-$ ($\times 10^{-16}$ W) | 1.49 | 1.88 | 2.29 | 2.72 | 3.17 | 3.63 | 4.11 | 4.60 |
| $I_-$ ($\times 10^{-11}$ A) | 8.38 | 8.58 | 8.77 | 8.97 | 9.17 | 9.36 | 9.57 | 9.79 |
| $R_-$ ($\times 10^4$ Ω) | 2.13 | 2.56 | 2.98 | 3.39 | 3.77 | 4.14 | 4.48 | 4.80 |
| $P_+$ ($\times 10^{-15}$ W) | 9.57 | 9.82 | 10.0 | 10.2 | 10.4 | 10.5 | 10.7 | 10.8 |
| $I_+$ ($\times 10^{-10}$ A) | 2.03 | 2.00 | 1.99 | 2.00 | 2.00 | 2.02 | 2.04 | 2.06 |
| $R_+$ ($\times 10^5$ Ω) | 2.33 | 2.45 | 2.53 | 2.57 | 2.58 | 2.58 | 2.57 | 2.54 |

Tab. 2. The radiation power ($P_\pm$), induced current ($I_\pm$) and radiation resistance ($R_\pm$) of bonding mode (with subscript -) and antibonding mode (with subscript +) under different coupling distance between two SRRs.

| D (nm) | 70 | 80 | 90 | 100 | 110 | 120 | 130 | 140 |
|---|---|---|---|---|---|---|---|---|
| $\omega_-$ (THz) | 59.02 | 60.60 | 61.91 | 63.01 | 63.93 | 64.71 | 65.38 | 65.94 |
| $\Delta\omega_-$ (THz) | 0.98 | 1.03 | 1.07 | 1.11 | 1.16 | 1.20 | 1.23 | 1.27 |
| $Q_-$ | 376.6 | 368.8 | 362.8 | 355.0 | 346.8 | 339.8 | 333.5 | 327.1 |
| $\omega_+$ (THz) | 93.12 | 91.95 | 90.79 | 89.67 | 88.60 | 87.58 | 86.62 | 85.71 |
| $\Delta\omega_+$ (THz) | 4.53 | 4.46 | 4.39 | 4.32 | 4.24 | 4.15 | 4.07 | 3.99 |
| $Q_+$ | 129.2 | 129.4 | 129.8 | 130.5 | 131.2 | 132.4 | 133.8 | 134.9 |

Tab. 3. The resonant frequency ($\omega_\pm$), FWHM ($\Delta\omega_\pm$) and Q factor ($Q_\pm$) at bonding (with subscript -) and antibonding mode (with subscript +) under different coupling distance between two SRRs (D).

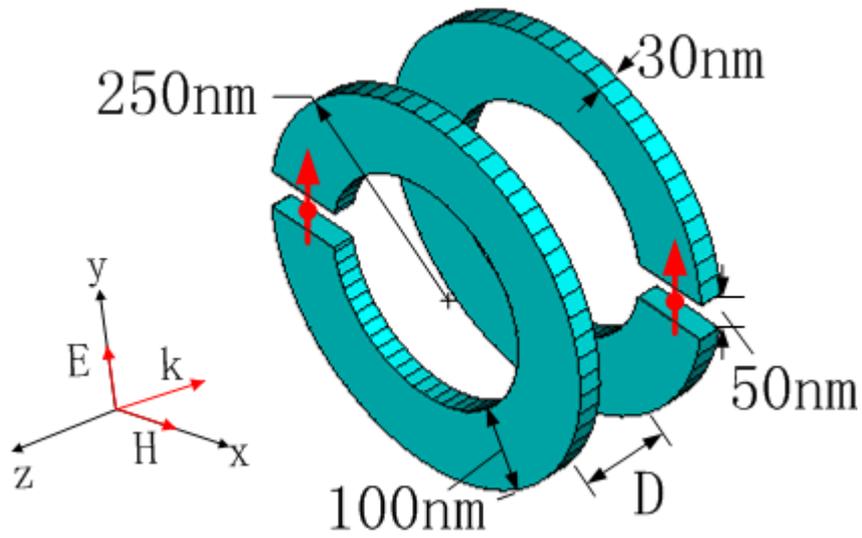

Figure 1. (Color online) Structure of the MD structure which is consisted of two spatially separated identical SRRs, twisted at an angle of $\varphi = 180^0$ with respect to one another. Two probes are set in the center of two splits to record the local field (the positions are showed as red arrows)

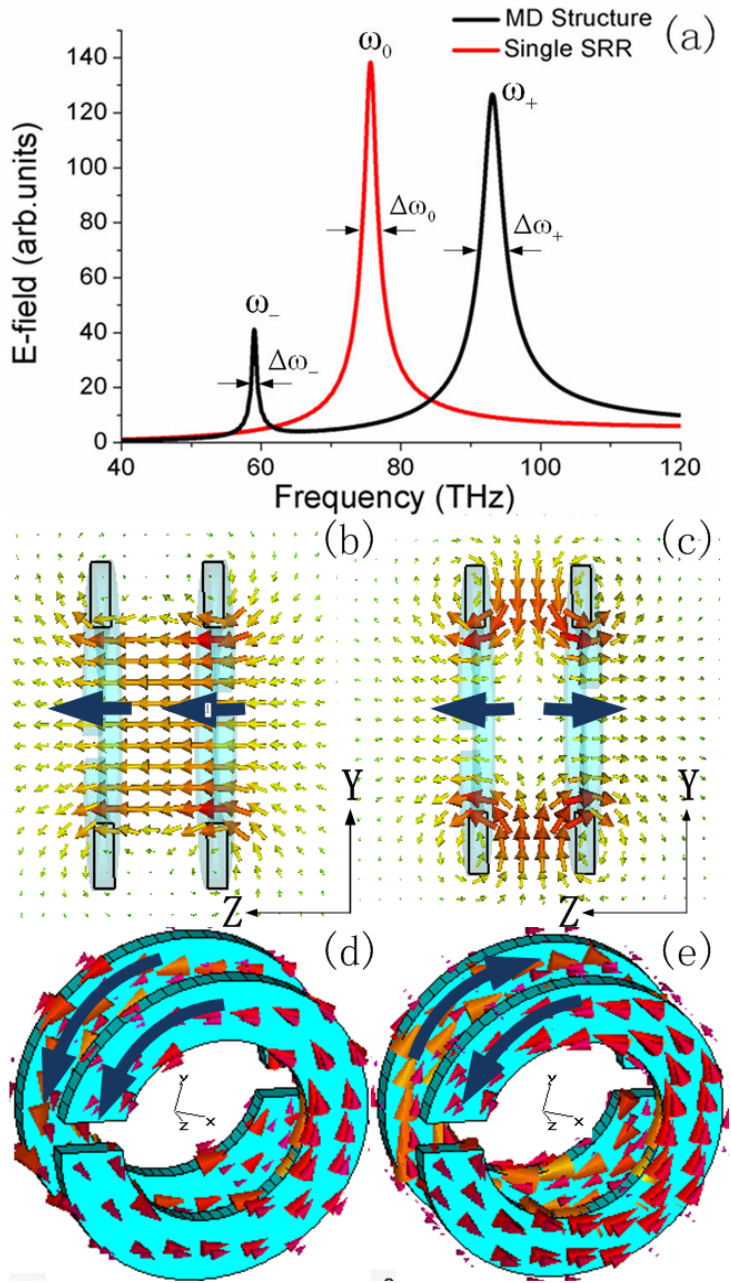

Figure 2. (Color online) (a) Local electric field detected by a probe in a single SRR structure (Red curve) and a MD structure (Black curve). Local magnetic field distribution in y-x plane ($z = 0$) at bonding mode (b) and antibonding mode (c). Local induced current distribution at bonding mode (d) and antibonding mode (e). The distance between two SRRs is set as $D = 70\text{nm}$.

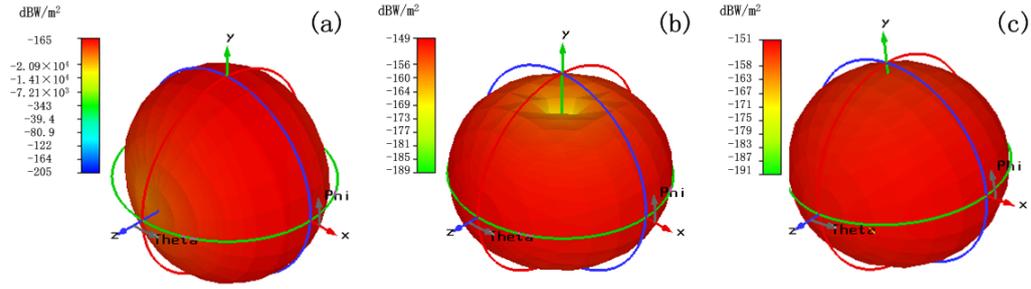

Figure 3. (Color online) The logarithmic radiation skin: (a) bonding mode of MD, (b) antibonding mode of MD, and (c) the single SRR. The distance between two SRRs is set as $D = 140\mathrm{nm}$.

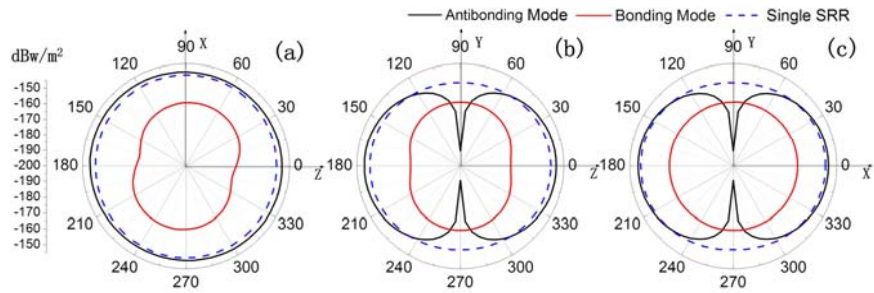

Figure 4. (Color online) The projection of the 3D radiation skin in (a) x-z plane ($y = 0$), (b) y-z plane ($x = 0$) and (c) x-y plane ($z = 0$). The distance between two SRRs is set as $D = 140\mathrm{nm}$.

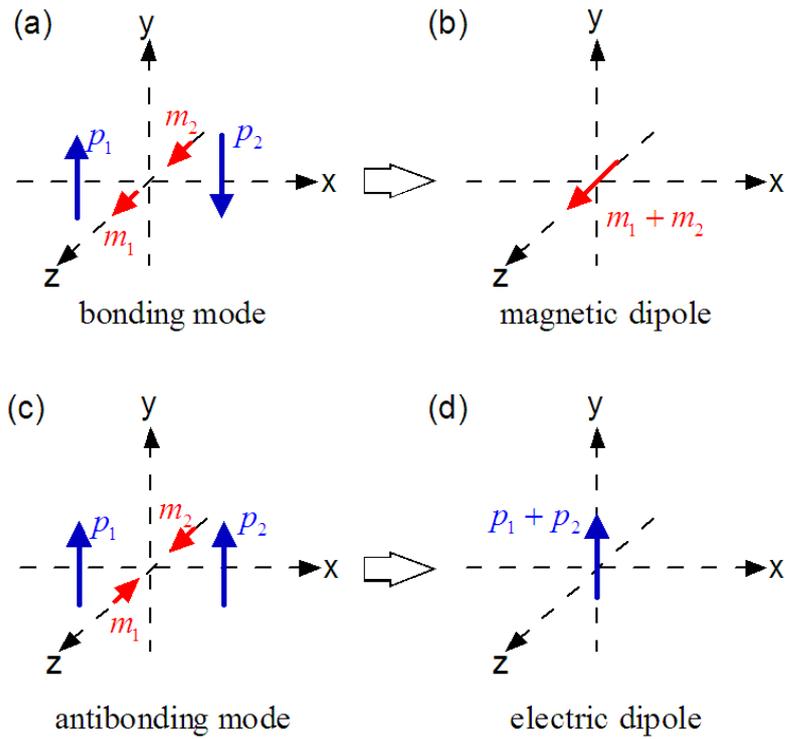

Figure 5. (Color online) The equivalent dipole model of MD at (a-b) bonding mode, (c-d) antibonding mode. (Red arrows: magnetic dipoles; blue arrows: electric dipoles)

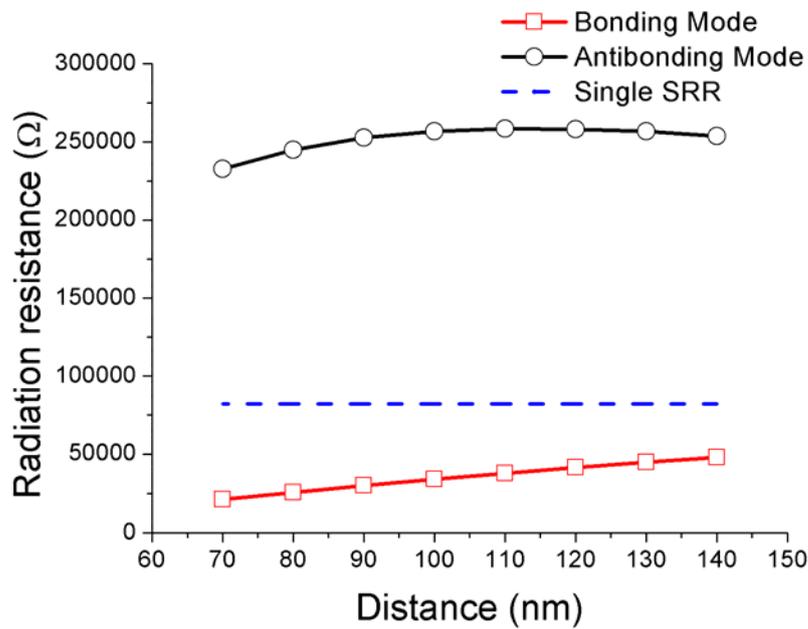

Figure 6. (Color online) Radiation resistance under different distances, red line: bonding mode of MD; black line: antibonding mode of MD; and blue dash line: the single SRR.

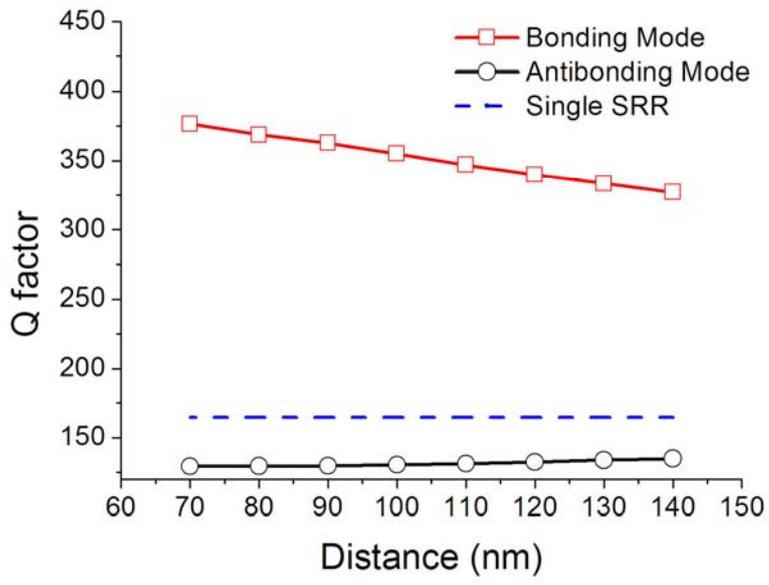

Figure 7. (Color online) Q-factors under different distances (red line: bonding mode of MD; black line: antibonding mode of MD; blue dash line: the single SRR)